\documentclass[twocolumn,aip,jcp,showpacs,floatfix]{revtex4}

\usepackage{epsf}
\usepackage[dvips]{graphicx}
\usepackage{bm}

\begin{document}

\title{
First principles investigation of finite--temperature behavior
in small sodium clusters}

\author {Mal-Soon Lee}   \email{mslee@unipune.ernet.in}
\author {S. Chacko}      \email{chacko@unipune.ernet.in}
\author {D. G. Kanhere}  \email{kanhere@unipune.ernet.in}

\affiliation{
Centre for Modeling and Simulation, and Department of Physics,
University of Pune, Ganeshkhind, Pune -- 411 007, India.}

\date{\today}

\begin{abstract}

A systematic and detailed investigation of the finite--temperature
behavior of small sodium clusters, Na$_n$, in the size range of $n=$ 8
to 50 are carried out.
The simulations are performed using density--functional
molecular--dynamics with ultrasoft pseudopotentials.
A number of thermodynamic indicators such as specific--heat, caloric
curve, root--mean--square bond--length fluctuation, deviation energy,
etc. are calculated for each of the clusters.
Size dependence of these indicators reveals several interesting
features.
The smallest clusters with $n=$ 8 and 10, do not show any signature of
melting transition.
With the increase in size, broad peak in the specific--heat is
developed, which alternately for larger clusters evolves into a
sharper one, indicating a solidlike to liquidlike transition.
The melting temperatures show irregular pattern similar to
experimentally observed one for larger clusters [ M. Schmidt {\it et
al.}, Nature (London) {\bf 393}, 238 (1998) ].
The present calculations also reveal a remarkable size--sensitive
effect in the size range of $n=$ 40 to 55.
While Na$_{40}$ and Na$_{55}$ show well developed peaks in the
specific--heat curve, Na$_{50}$ cluster exhibits a rather broad peak,
indicating a poorly--defined melting transition.
Such a feature has been experimentally observed for gallium and
aluminum clusters [ G. A. Breaux {\it et al.}, J. Am. Chem. Soc.,
{\bf 126}, 8628 (2004); G. A.Breaux {\it et al.}, Phys. Rev. Lett.,
{\bf 94}, 173401 (2005) ].

\end{abstract}

\maketitle

\section{Introduction \label{intro}}

Finite temperature studies of finite--sized systems have been a topic
of considerable interest during last decades.
Recent experimental as well as theoretical studies have brought out a
number of intriguing results.
In a series of experiments on free sodium clusters in the
size range of 55 to 350, Haberland and co--workers~\cite{haberland}
have observed a large size--dependent fluctuation in the melting
temperatures.
They have also observed a substantial lowering of about 30~\% in the
melting temperatures as compared to that of the bulk.
Interestingly, recent experiments by Jarrold and
coworkers~\cite{jarrold-sn, jarrold-ga1} show that small clusters of
Sn and Ga have higher--than--bulk melting temperatures.
In our previous investigations, we have attributed this
higher--than--bulk melting temperatures mainly to the covalent bonding
in these clusters, as against metallic bonding in the bulk
phase.~\cite{our-sn, our-ga}
Very recently Breaux {\it et al.} have seen a remarkable
size--sensitive feature of the melting transition of gallium as well
as aluminum clusters.~\cite{jarrold-ga2, jarrold-al}
Their experiments show that the nature of the heat capacity curve
changes dramatically with addition of few atoms.
For instance, Ga$_{30}^+$ does not show obvious melting transition,
while Ga$_{31}^+$ exhibit a well--defined peak, and Ga$_{32}^+$ shows
a broad peak in the heat capacity.

A number of computer simulations on melting of small sodium clusters
have been reported in literature.
Calvo and Spiegelmann~\cite{Calvo} have performed extensive
simulations on Na clusters in size range of 8 to 147.
Their simulations employed the second moment approximation~(SMA)
potential of Li {\it et al.}~\cite{SMA} as well as the
distance--dependent tight--binding (DDTB or TB) method.
They found more than one peak in the heat capacity most of of the
clusters studied.
They further observed that the nature of the ground--state geometry
is crucial to precisely understand the thermodynamic properties of
clusters.
However, although the method they employed provide relatively good
statistics required to converge the features in the caloric curve,
it did not incorporate the essential ingredients of electronic
structure effects.
These simulations hence failed to reproduce the crucial features of
the experimental results, clearly bringing out the importance of
incorporating the electronic structures effects.
In a very recent study~\cite{Na55-142}, we have successfully
reproduced the melting temperatures of Na$_N$ ($N=55, 92, 142$) using
the Kohn--Sham (KS) based approach~\cite{KS} of the
density--functional theory (DFT) and also gave a plausible explanation
on its irregular variation.
There have also been a few theoretical investigations on melting of
sodium clusters with sizes $n<55$.
Manninen {\it et al.}~\cite{Manninen} have investigated the melting
transition of Na$_{40}$ cluster using {\it ab initio} method.
They raised the temperature of the system from 150~K to 400~K at the
rate of 5~K/$ps$, and found that the melting transition occurs at the
temperatures between $300\sim350$~K Na$_{40}$.
They also show that Na$_8$ exhibits only isomerization.
Aguado {\it et al.}~\cite{Aguado} have performed Car--Parrinello
orbital--free simulations to investigate melting phenomena in Na$_8$
and Na$_{20}$.
Their simulation times were 8--60~$ps$ per temperature.
They observed a clear peak for Na$_8$ and double peaks for Na$_{20}$
in the specific--heat curve.
We have also investigated the melting transition in these clusters
using various methods, and found a model dependence in the melting
characteristics.~\cite{Na-AMV}

In the present work, we perform density--functional
molecular--dynamical simulations on Na$_n$ clusters ($n=$ 8, 10, 13,
15, 20, 25, 40, and 50) to investigate their thermodynamic properties,
specifically the size--dependent features.
We perform simulations with about 150~$ps$ per temperature which is
much larger simulation times than any other earlier work.
In addition to the standard indicators like specific--heat, caloric
curve, root--mean--square bond--length fluctuations, we also calculate
the energy deviation, the potential energy difference between the
solidlike--state and the liquidlike--state, etc.

In Sec.~\ref{comp}, we describe the computational details, followed
by the results and discussion in Sec.~\ref{results}.
Finally, we summarize the results in Sec.~\ref{summ}.

\section{Computational Details \label{comp}}

We carry isokinetic Born--Oppenheimer molecular--dynamics
calculations~\cite{BOMD} using Vanderbilt's ultrasoft
pseudopotentials~\cite{uppot} within the local--density
approximation~(LDA), as implemented in the VASP package.~\cite{VASP}
We use two different methods to obtain the ground--state and several
equilibrium geometries for each of the clusters.
First, a ``basin hopping'' algorithm~\cite{basin} is employed to
generate few tens of structures for smaller clusters and several
hundreds structures for larger clusters using the second moment
approximation~(SMA) parameterized potential of Li {\it et
al.}~\cite{SMA}
Several of these geometries, say the lowest 10--70 geometries, are
then optimized using {\it ab initio} density--functional
method.~\cite{DFT}
In the second method, we obtained few more equilibrium geometries by
optimizing several structures selected from high--temperature {\it ab
initio} molecular--dynamics runs, typically taken from temperatures
near and well above the melting temperatures of the clusters.
The simulations have been carried out for 12 temperatures in the range
of $100K\le T\le750K$ for $n=$ 8 and 10, 9--12 temperatures in the
range of $100K\le T\le450K$ for the rest clusters.
For all the cases, the simulation time is 150~$ps$ per temperature.
We have discarded first 30~$ps$ for each temperature to allow
for thermalization.
An energy cutoff of 3.6~$Ry$~\cite{encut-conv} is used for the
plane wave expansion of the wavefunction, with a convergence in the
total energy of the order of 10$^{-4}$~eV.
The resulting ionic trajectory data have been used to study the
melting of clusters by analyzing various thermodynamic indicators,
which are discussed below in detail.

We calculate the deformation parameter, $\varepsilon_{def}$, to
analyze the shape of the ground--state geometry for all the clusters.
The shape of the ground--state geometry plays a crucial role in
determining the thermodynamic properties of a cluster.
The deformation parameter, $\varepsilon_{def}$, is defined as
$$
\varepsilon_{def} = {2Q_1 \over Q_2+Q_3}
$$
where $Q_1 \ge Q_2 \ge Q_3$ are eigenvalues of the quadrupole tensor
$
Q_{ij} = \sum_I R_{Ii}R_{Ij}
$
\noindent
with $R_{Ii}$ being {\it i}$^th$ coordinate of ion $I$ relative to the
center of mass of the cluster.
A spherical system ($Q_1 = Q_2 = Q_3$) has $\varepsilon_{def}$ = 1,
while $\varepsilon_{def} > $ 1 indicates a quadrupole deformation of
some kind.

To analyze the thermodynamic properties, we first calculate the ionic
specific--heat and the average potential energy per temperature
(the caloric curve).
We extracted the classical ionic density of states, $\Omega(E)$,
of the system, or equivalently the classical ionic entropy,
$S(E)=k_B \ln \Omega(E)$, via the multiple histogram
method~\cite{MH-method} to evaluate the canonical specific--heat.
In the canonical ensemble, the specific--heat is defined as
$C(T)=\partial U(T)/\partial T$, where $U(T)=\int{E~p(E,T)dE}$ is the
average total energy.
The probability of observing an energy $E$ at a temperature $T$ is
given by the Gibbs distribution $p(E,T)=\Omega(E) \exp(-E/k_BT)/Z(T)$,
with $Z(T)$ the normalizing canonical partition function.
We normalize the calculated canonical specific--heat by the zero
temperature classical limit of the rotational plus vibrational
specific--heat, {\it i.e.} $C_0=(3N-9/2)k_B$.
Details of this method can be found in Ref.~\onlinecite{Na-AMV}.
The melting temperature has been taken as a peak in the specific--heat
curve, following the convention of the experiments.~\cite{haberland}

Other thermodynamic indicator calculated is the root--mean--square
bond length fluctuations~(RMSBLF), {\it i.e} the Lindemann--like
criterion for a finite system, given as
$$
\delta_{rms} = {2 \over  N(N-1)}
               \sum_{i<j}
               {\sqrt{\langle R_{ij}^{2} \rangle_{t}-
                      \langle R_{ij}     \rangle_{t}^{2}}
                  \over
                \langle R_{ij}\rangle_{t}}
$$
where, $R_{ij}$ is the distance between $i^{th}$ and $j^{th}$ ion.
This quantity gives the average fluctuation in the average
bond lengths that are occurring at a given temperature.
A value of about 0.1--0.15 signifies a melting transition for the
bulk.
However, as we shall see, for smaller clusters, this indicator should
be taken with some caution.
It is useful when examined in conjunction with other indicators such
as the specific--heat.

We have also calculated the energy deviation, $\delta
E$,~\cite{Chelikowsky} defined as
$$
\delta E=\langle  E_{total}(T) \rangle-[E_0+(3n-6)k_BT]
$$
where, $\langle  E_{total}(T) \rangle=\langle  E_{kin}(T) \rangle+
\langle  E_{pot}(T) \rangle$ is the average total energy of the system
at temperature $T$, with $\langle E_{kin}(T)\rangle$ and $\langle
E_{pot}(T) \rangle$ being the average kinetic energy and the average
potential energy, respectively.
$E_0$ is the ground--state energy and $(3n-6)k_BT$ is the vibrational
energy in the classical limit.
$\delta E$ may be considered to be an indicator of anharmonicity in
the system, as a function of temperature.
Following Chuang {\it et al.}~\cite{Chelikowsky}, we smoothened the
plots of $\delta E$ using three point moving average method.
The error bars in the plot are the standard errors of every three
data points.

\begin{figure}
  \includegraphics[width=3.0in]{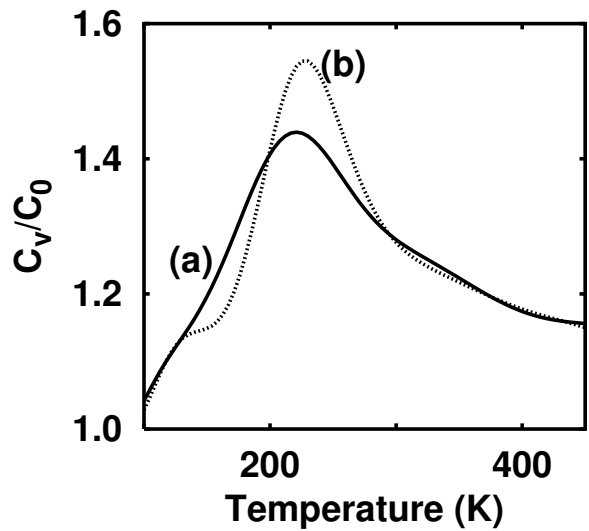}
  \caption{
   The specific--heat of the Na$_{20}$ cluster with simulation time of
   (a) 150$~ps$ and (b) 90$~ps$ as a function of temperature.
   $C_0=(3N-9/2)k_B$ is the zero--temperature classical
   limit of the rotational plus vibrational canonical
   specific--heat.}
  \label{fig.na20.cv.60ps.vs.120ps}
\end{figure}

We have also examined carefully the role of simulation time.
For this purpose, in Fig.~\ref{fig.na20.cv.60ps.vs.120ps}, we plot the
specific--heat for Na$_{20}$ with two simulation times: one with
90~$ps$, and another with 150~$ps$, per temperature.
It may be immediately seen that the 90~$ps$ data results in a
premelting feature which is absent for the 150~$ps$ one.
This indicates that even for smaller systems such as this, one needs
to go to higher simulation times of the order of 150~$ps$ or so.

\section{Results and Discussion \label{results}}

We begin by analyzing the geometries of sodium clusters for the sizes
of $n=$ 8, 10, 13, 15, 20, 25, 40 and 50.
This is then followed by a discussion on their thermodynamic
properties.
We also address certain features of Na$_{55}$ and Na$_{92}$ clusters
relevant to the present discussion.~\cite{Na55-142}

\subsection{Geometry}

\begin{figure}
  \includegraphics[width=3.4in]{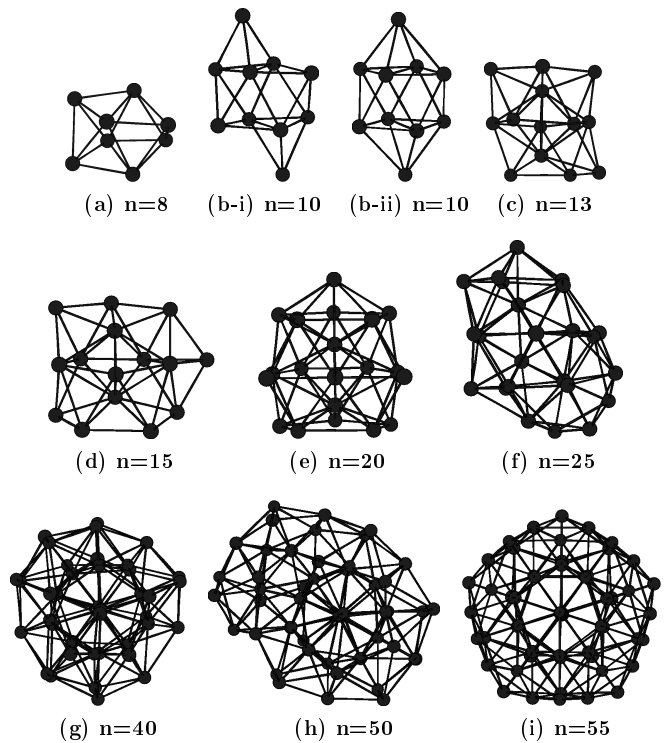}
  \caption{
   The ground--state geometries of the Na$_n$ ($n=8-55$) clusters.}
  \label{fig.gs_geom}
\end{figure}

\begin{figure}
  \includegraphics[width=2.8in]{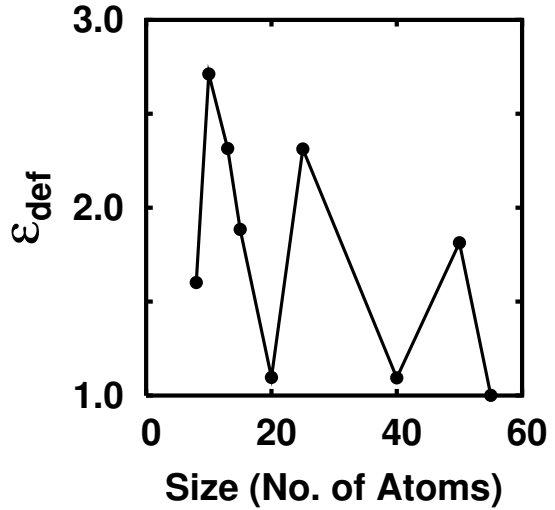}
  \caption{
   The deformation parameter, $\varepsilon_{def}$, of the
   ground--state geometries as a function of cluster size.}
  \label{fig.GS.eps_pro}
\end{figure}

\begin{figure}
  \includegraphics[width=3.4in]{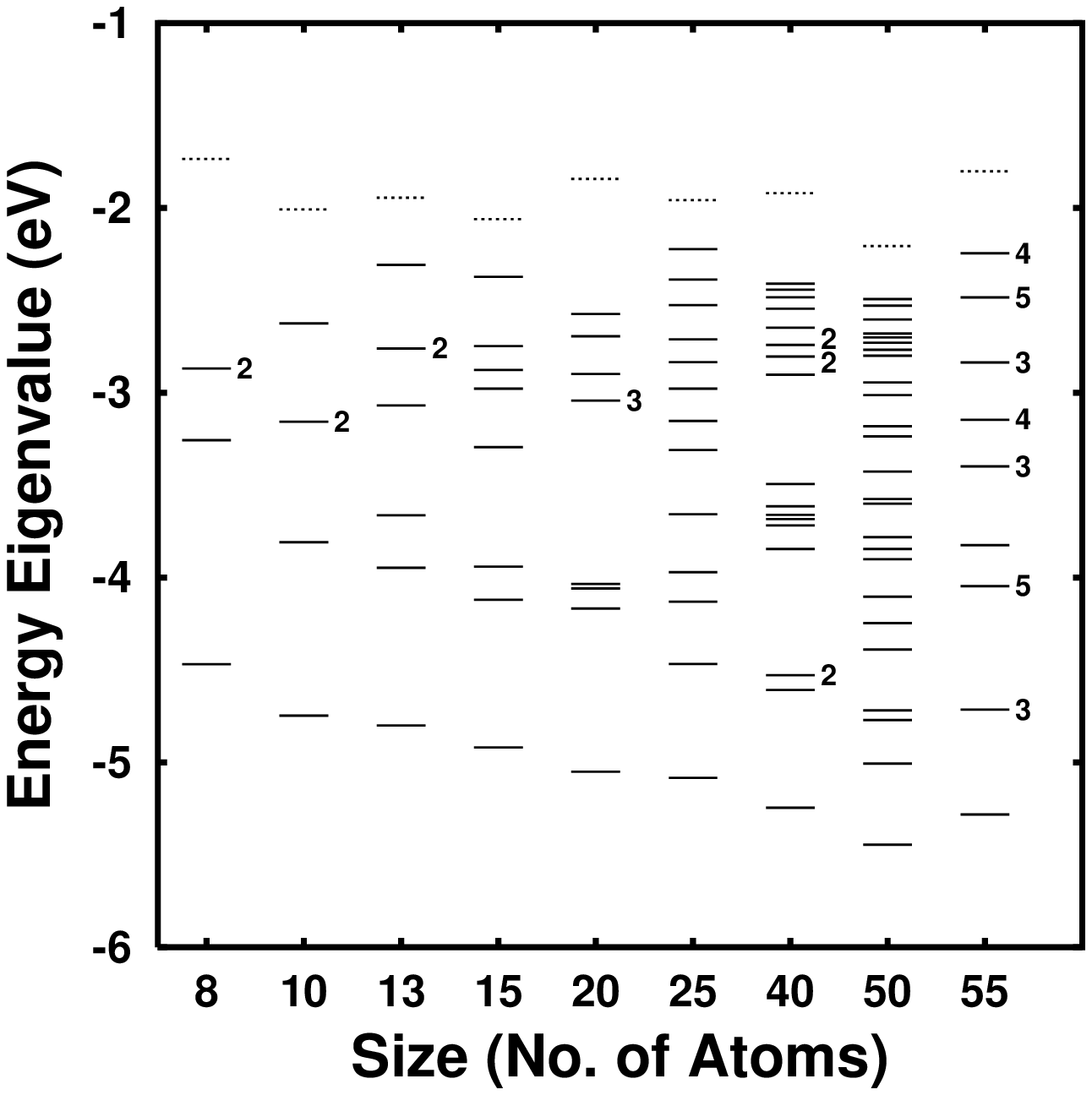}
  \caption{
   The eigenvalue spectra of their ground--state of Na$_n$ ($n-8-50$)
   cluster as a function cluster size.
   The numbers on the right show the degeneracy of that level.}
  \label{fig.GS.evs}
\end{figure}

The lowest--energy geometries of sodium clusters are shown in
Fig.~\ref{fig.gs_geom}.
First, we note that the equilibrium geometries of Na$_8$, Na$_{13}$
and Na$_{20}$, obtained by us, are in agreement with those reported
by R\"othlisberger \emph{et al.}~\cite{Ursula}
The ground--state geometry of Na$_8$ (Fig.~\ref{fig.gs_geom}(a)) is
a dodecahedron.
One of its low--energy isomer is an antiprism ($\Delta~E=0.04$~eV).
These two structures play a crucial role in the finite temperature
behavior of this cluster.
We find two nearly degenerate structures for Na$_{10}$, namely a
bicapped dodecahedron (Fig.~\ref{fig.gs_geom}(b--i)) and a bicapped
antiprism (Fig.~\ref{fig.gs_geom}(b--ii)).
R\"othlisberger \emph{et al.}~\cite{Ursula} have found the bicapped
dodecahedron to be unstable.
However, we computed the vibrational spectra for both geometries and
found the structures to be stable.
The clusters, Na$_{13}$ (Fig.~\ref{fig.gs_geom}(c)) and Na$_{15}$
(Fig.~\ref{fig.gs_geom}(d)), exhibit capped pentagonal bipyramidal
structures as their lowest--energy configurations.
The ground--state geometry of Na$_{20}$ (Fig.~\ref{fig.gs_geom}(e))
is a fivefold capped icosahedron, with two atoms capping the
icosahedral faces on the central plane.
We find a capped double icosahedron to be about 0.11~eV higher in
energy than the ground--state.
The lowest--energy structure of Na$_{25}$ has not been previously
reported.
It may be described as capped double icosahedron with growth on one
side, leading to a distorted non--spherical structure, as shown in
Fig.~\ref{fig.gs_geom}(f).
The ground--state geometry for Na$_{40}$, shown in
Fig.~\ref{fig.gs_geom}(g), agrees with the one reported by Manninen
\emph{et al.}~\cite{Manninen}
It consists of three decahedra capped by rest of the atoms.
The structure is compact and retains the five--fold symmetry.
The ground--state geometry of Na$_{50}$, shown in
Fig.~\ref{fig.gs_geom}(h), is highly asymmetric.
It can be seen from figure~\ref{fig.gs_geom} that a growth towards
50--atom structure starting from symmetric Na$_{40}$ makes the
structure non--spherical and asymmetric, which is similar as seen for
Na$_{25}$ cluster.
Finally, the ground--state geometry of Na$_{55}$ is a slightly
distorted double Mackay icosahedron.
It is the most spherical structure and is noted for the sake of
completeness.~\cite{Na55-142}

We have also examined the eigenvalue spectra and the shapes of the
ground--state geometries of these clusters, as shown in
Figs.~\ref{fig.GS.evs} and \ref{fig.GS.eps_pro}, respectively.
The shape deformation parameter, $\varepsilon_{def}$, plotted in
Fig.~\ref{fig.GS.eps_pro}, for the ground--state geometries of all the
clusters show that Na$_{20}$, Na$_{40}$ and Na$_{55}$ are nearly
spherical, while Na$_{25}$ and Na$_{50}$ are deformed.
This behavior is also reflected in the eigenvalue spectra of these
clusters.
The eigenvalue spectra of Na$_{20}$, Na$_{40}$ and Na$_{55}$ clusters
(see Fig.~\ref{fig.GS.evs}), having nearly spherical geometries,
conforms the jellium description.
For instance, Na$_{55}$ shows a jellium--like behavior with {\it s, p,
d, ...} shell structure.
However, for systems such as  Na$_{25}$ and Na$_{50}$, due to the
disordered nature of the ground--state geometries, the degeneracy in
the eigenvalue spectra is lifted, thereby leading to a continuous
spectra.

\begin{figure}
  \includegraphics[width=3.4in]{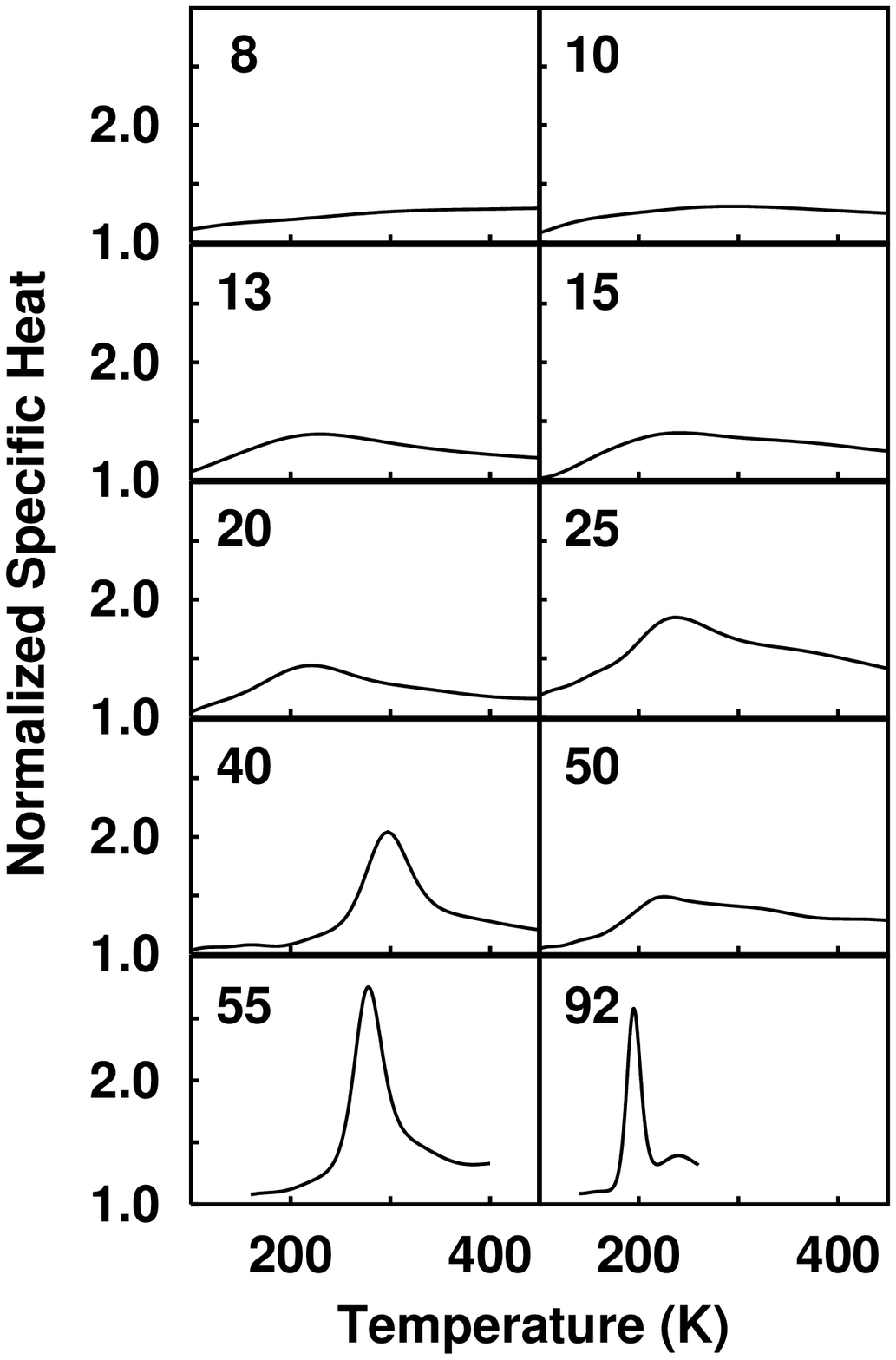}
  \caption{
   The normalized specific--heat as a function of temperature.
   $C_0=(3N-9/2)k_B$ is the zero--temperature classical
   limit of the rotational plus vibrational canonical
   specific--heat}.
  \label{fig.cv}
\end{figure}

\subsection{Thermodynamics}

\begin{figure}
  \includegraphics[width=3.4in]{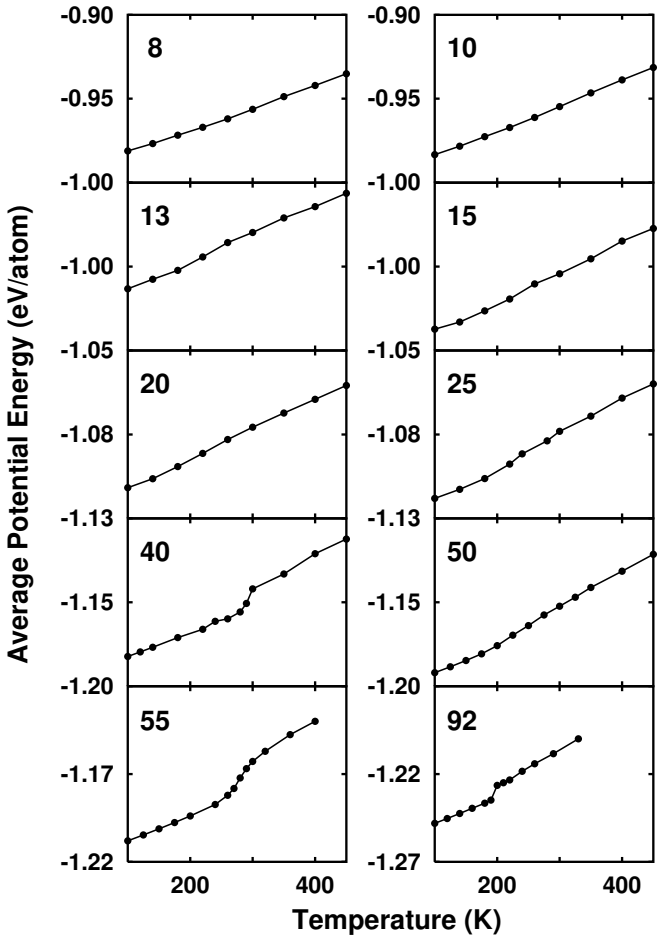}
  \caption{
   The averaged potential energy (caloric curve) as a function of
   temperature.}
  \label{fig.caloric}
\end{figure}

\begin{figure}
  \includegraphics[width=3.2in]{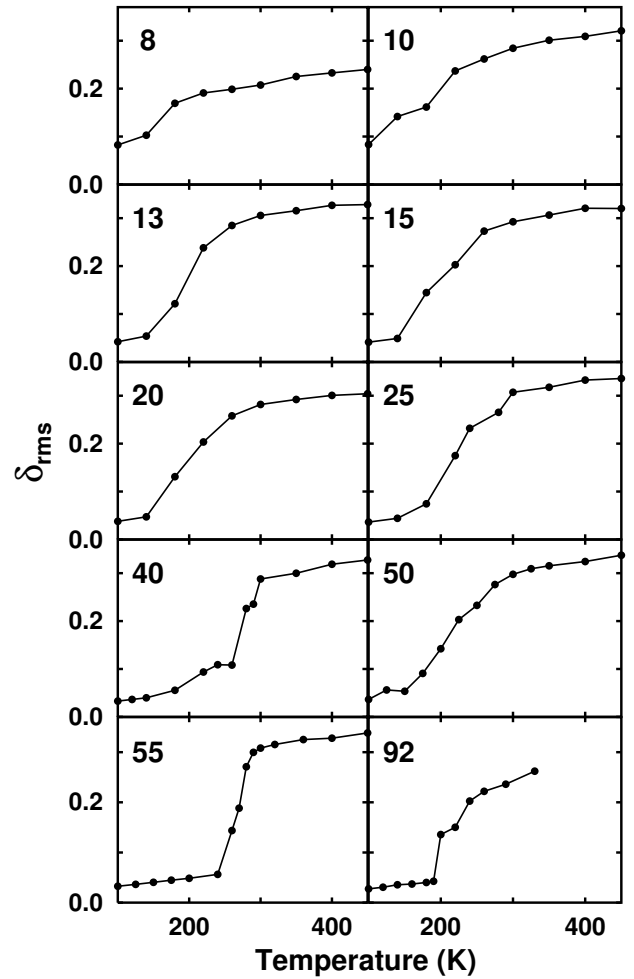}
  \caption{
   The root--mean--square bond length fluctuation ($\delta_{rms}$) as a
   function of temperature.}
  \label{fig.delta}
\end{figure}

\begin{figure}
  \includegraphics[width=2.8in]{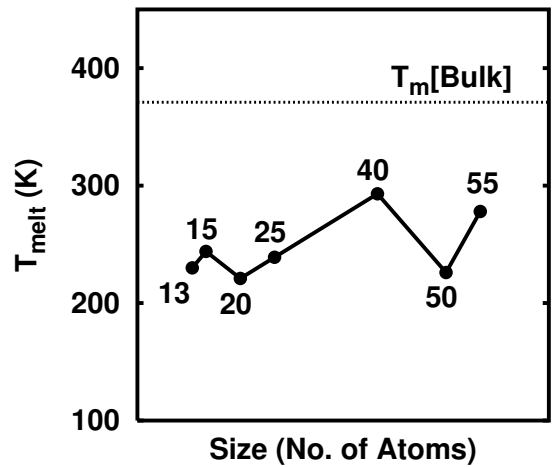}
  \caption{
   The melting temperature as a function of size.
   The melting temperature in bulk (T$_m$[Bulk]) is 370~K is shown
   with the horizontal line.
   }
 \label{fig.tm}
\end{figure}

The thermodynamic behavior is studied by analyzing several indicators.
We have calculated the specific--heat, caloric curve, Lindemann
criterion ($\delta_{rms}$) as a function of temperature for each
cluster.
These are shown in Figs.~\ref{fig.cv}, \ref{fig.caloric} and
\ref{fig.delta}, respectively.
The melting temperatures, $T_m$, taken as the temperature
corresponding to the peak in the specific--heat curve, are shown in
Fig.~\ref{fig.tm}.

\begin{figure}
  \includegraphics[width=3.4in]{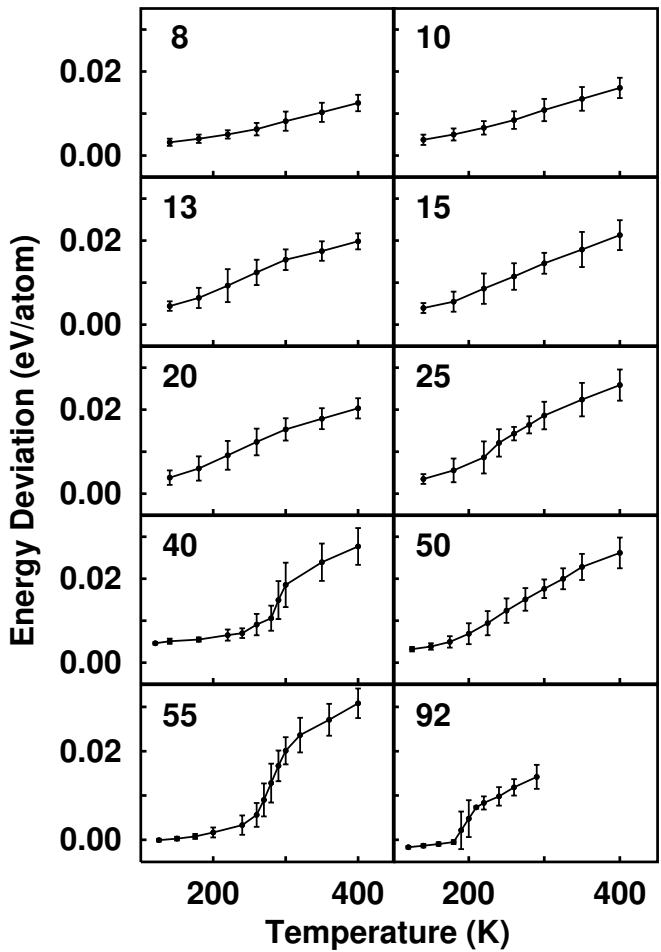}
  \caption{
   The deviation of total energy from that of the harmonic limit
   plotted as a function of temperature.
   The error bar is the standard error at each temperature.}
  \label{fig.eng_dev}
\end{figure}

The examination of thermodynamic indicators as a function of cluster
size reveals interesting trends.
It may be seen that the clusters of sizes 8 and 10, do not show any
recognizable peak in the heat capacities.
This is also reflected in the caloric curves which increase
continuously.
While the peak is rather broad for $n=$ 13--20, it progressively
becomes narrower as the size increases, and at $n=$ 92, a rather sharp
peak with a width of the order of 30~K is observed.
The caloric curve (Fig.~\ref{fig.caloric}) and the $\delta_{rms}$
(Fig.~\ref{fig.delta}) for larger clusters, viz. $n=$ 40, 55, and 92,
clearly show distinct solidlike, liquidlike and transition regions.
Interestingly, the melting temperatures depicted in Fig.~\ref{fig.tm}
show irregular pattern with the maximum variation of about 60~K.
High melting temperatures are observed in two clusters: Na$_{40}$ and
Na$_{55}$, which are very symmetric; Na$_{40}$ exhibiting electronic
closure and Na$_{55}$ representing a geometrically closed system.
We plot $\delta E$ in Fig.~\ref{fig.eng_dev}, which is an indicator of
anharmonicity in the system.
It is clear from the figure that for the clusters with $n\ge$ 40, it
is possible to distinguish a temperature region ($<$200~K) showing
harmonic behavior.
In contrast to this, the smaller clusters change from harmonic to
anharmonic behavior nearly continuously.
The most remarkable observation concerns the trends in the
specific--heat curve and other indicators for sizes $n=$ 40, 50 and
55.
In spite of a well--defined peak in the specific--heat curves for
Na$_{40}$ and Na$_{55}$, the Na$_{50}$ cluster shows a rather broad
structure, similar to that seen in smaller clusters.
This size--sensitive behavior is discussed further below.

\begin{figure}
  \includegraphics[width=3.0in]{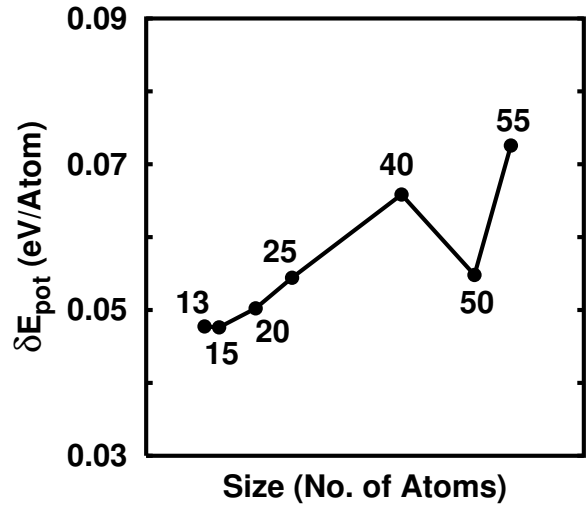}
  \caption{
   The difference of average potential energy $\delta E_{pot}$ between
   solid--state and liquid--state as a function of size.}
  \label{fig.dE_pot}
\end{figure}

Thus, all the indicators, like the specific--heat, the caloric curve,
$\delta_{rms}$ and $\delta E$ clearly show that small clusters
($n=$ 8, 10) do not undergo any melting--like transition.
The examination of their ionic motion indicates that over the entire
range of the temperatures the motion is dominated by isomer hopping.
This is in agreement with the observation by Manninen {\it et
al.}~\cite{Manninen}
However, these results are in contrast with the SMA and tight binding
calculations of Calvo {\it et al.}~\cite{Calvo}, and the
Car--Parrinello orbital free calculations of Aguado {\it et
al.}~\cite{Aguado}
Recall that these calculations, though are not in agreement with each
other, show distinct peaks (broader in case of SMA) in the heat
capacities of these clusters.
Our simulations further show the smaller clusters to be dissociated at
about 750~K.
Our results for Na$_{13}$ and Na$_{20}$ are also in disagreement
with those by Calvo {\it et al.}~\cite{Calvo}
They calculated the canonical heat capacity of Na$_{13}$, using an
icosahedron for SMA calculations and a pentagonal structure with $C_1$
symmetry for TB calculations as the ground--state geometries.
While the heat capacity with the SMA potential exhibited a single
prominent peak, that of TB calculation showed a premelting feature.
They attributed this difference to the difference in the ground--state
geometries.
They further found such premelting feature in the heat capacity of
Na$_{20}$ cluster, for which they used a capped double icosahedron as
the lowest energy structure.
However, our calculations show that this structure is about 0.1~eV
higher than the ground--state structure obtained by us for Na$_{20}$.
Thus, the differences in the specific--heat may be also caused by the
differences in the ground--state geometries.
The $\delta_{rms}$ for all the clusters (Fig.~\ref{fig.delta}) clearly
shows that for smaller systems ($n<$ 40), it increases almost
continuously, whereas for larger ones, a sharp rise is seen that
corresponds to the peak in the specific--heat.
The behavior of $\delta E$ is, as noted earlier, consistent with this.
We have also calculated a quantity $\delta E_{pot}$ defined as the
difference of average potential energy of the melted cluster with
respect to the ground--state structure at $T=0$~K.
Schmidt \emph{et al.}\cite{schmidt-na-dE} have inferred from the
experimental caloric curve that the melting temperature is strongly
influenced by such an energy contribution.
They showed further that $T_m$ follows closely the variation in
energy difference between solid and liquid as a function of the
cluster size.
To examine this feature, we plot the potential energy difference
between solidlike--state and liquidlike--state shown in
Fig.~\ref{fig.dE_pot}.
For this purpose, we have taken the high temperature as 300~K for $n=$
13, 15, 20, 25, and 50, 350~K for $n=$ 40, 55, which are about 50~K to
70~K higher than the melting temperature of the respective clusters.
It can be immediately seen that the variation of $\delta E_{pot}$
follows that of the melting temperature (Fig.~\ref{fig.tm}), clearly
indicating that the melting transition is mainly driven by energy
contribution.

\subsection{Na$_{40}$ $\cdot$ Na$_{50}$ $\cdot$ Na$_{55}$}

Now, we turn our discussion to the most remarkable observation
concerning the trends in the specific--heat in going from $n=$ 40 to
$n=$ 55.
As mentioned earlier, while Na$_{40}$ and Na$_{55}$ show well--defined
peaks in the specific--heat curve (peak for Na$_{55}$ being much
sharper), the peak for Na$_{50}$ is rather flat and is almost similar
to that of the smaller clusters (say, $n=$ 13--20).
We note that Na$_{50}$, being larger than Na$_{40}$, is expected to
show slightly better melting transition.
Interestingly, what is seen is exactly the opposite.
It may be noted that such peculiar size--sensitivity has been observed
experimentally in two systems, namely clusters of gallium (Ga$_n^+$,
$n=$ 30--50 and 55)~\cite{jarrold-ga2} and clusters of aluminum
(Al$_n^+$, $n=$ 49--62)~\cite{jarrold-al}.
For instance, in case of Ga clusters the heat capacity for $n=$ 30
shows a flat curve without a peak.
For $n=$ 31 it shows a remarkably sharp peak.
Interestingly, addition of one more atom ({\it i.e.} $n=$ 32)
diminishes this peak making the heat capacity nearly flat.
We believe this behavior to be generic as it has not only been
observed experimentally in case of gallium clusters but also for
aluminum cluster and theoretically for sodium clusters in the present
simulations.

We note certain peculiar characteristics of the thermodynamic
properties of Na$_{40}$, Na$_{50}$, and Na$_{55}$.
We find the melting temperature of Na$_{50}$ to be about 60~K lower
that that of Na$_{40}$ and Na$_{55}$.
The $\delta_{rms}$ for Na$_{50}$ exhibits gradual increase in the
temperature range of 100~K to 300~K.
Further, the energy deviation $\delta E$ for Na$_{50}$, as seen in
Fig.~\ref{fig.eng_dev}, starts to increase continuously at about 100~K
to up to about 400~K, indicating a continuous change from harmonic
behavior to anharmonic one.
This behavior of Na$_{50}$ is in contrast with that of Na$_{40}$ and
Na$_{55}$, where the change is seen in a smaller temperature width (of
about 30--40~K) around the melting temperature.
In order to bring out the origin of this phenomena, we examine the
nature of the ground--state geometries for these three clusters.
Na$_{40}$ and Na$_{55}$ are very symmetric structures having almost
five--fold symmetry.
The values of the shape deformation parameter, $\varepsilon_{def}$,
shown in Fig.~\ref{fig.GS.eps_pro}, clearly indicate that they are
nearly spherical structures.
Further, the eigenvalue spectra of the ground--state geometries of
Na$_{40}$ and Na$_{55}$ clusters also show this symmetry, conforming
the jellium model.
However, the eigenvalue spectrum of Na$_{50}$ is very different from
those of Na$_{40}$ and Na$_{55}$, in the sense that there are levels
in the energy gaps leading to a more uniform spectrum.
In this sense, Na$_{40}$ and Na$_{55}$ are ordered, {\it i.e.} more
symmetric, and Na$_{50}$ is amorphous.

\begin{figure}
  \includegraphics[width=3.2in]{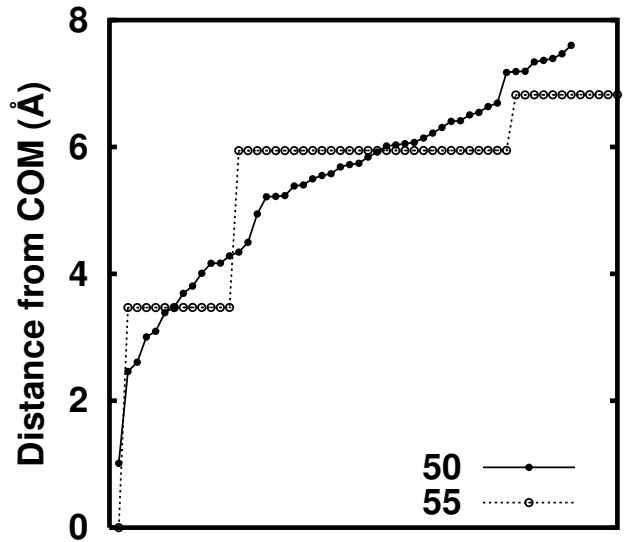}
  \caption{
   The distance from the center of mass for the ground--state
   geometries of Na$_{50}$ and Na$_{55}$.}
  \label{fig.dist_from_com}
\end{figure}

In Fig.~\ref{fig.dist_from_com}, we show the distances from the center
of mass of all the atoms in the ground--state geometries of Na$_{50}$
and Na$_{55}$ .
Clearly, the ordered geometric shell structure of Na$_{55}$ is
destroyed when five atoms are removed.
We believe that the nature of ground--state geometry of the cluster
has a significant effect on its melting characteristics.
An ordered or symmetric cluster, like Na$_{40}$ and Na$_{55}$, is
expected to give rise to a well--defined peak in the heat capacity,
while an amorphous and disordered clusters, like Na$_{25}$ and
Na$_{50}$, may lead to a continuous melting transition.

\section{Summary \label{summ}}

We have investigated the thermodynamics properties of small sodium
clusters, Na$_{n}$, in the size range of 8 to 55, using {\it ab
initio} molecular dynamics with simulation time of 1.3--1.8~$ns$ per
cluster.
We have analyzed several thermodynamic indicators such as the
specific--heat, caloric curve, Lindemann criterion, and the deviation
energy $\delta E$ to understand the melting characteristics in these
clusters.
We observe irregular variation in the melting temperatures as a
function of size, which has also been seen in experiments by Haberland
and coworkers for larger clusters.
The reduction of about 30~\% than the bulk value in melting
temperature of sodium clusters, seen in experiment is also observed
here.
Further, we find a strong correlation between the ground--state
geometry and the finite temperature characteristics of the sodium
clusters.
If a cluster has an \emph{ordered} geometry, it is likely to show a
relatively sharp melting transition.
However, a cluster having a \emph{disordered} geometry is expected to
exhibit a broad peak in the specific--heat curve, indicating a
poorly--defined melting transition.
The size--sensitivity in the melting transition, seen in experiments
by Breaux {\it et al.}~\cite{jarrold-ga2, jarrold-al}, is observed for
the case of sodium clusters in the size range of 40 to 50.
The calculation of potential energy difference between solidlike state
and liquidlike state reveals that the melting transition is mainly
driven by the energy contribution and the entropy has a minor role in
melting phenomena.

\section{Acknowledgments}

It is a pleasure to acknowledge C--DAC (Pune) for the supercomputing
facilities.
We also acknowledge partial assistance from the Indo--French Center by
providing the computational support.
One of us (SC) acknowledges financial support from the Center for
Modeling and Simulation, University of Pune and the Indo--French
Center for Promotion for Advance Research (IFCPAR).
We would like to thank Sailaja Krishnamurty for a number of
useful discussions.


\begin{thebibliography}{000}
\bibitem{encut-conv}
   In our earlier work~(see Ref.~\onlinecite{Na55-142}), we have
   verified that an energy cutoff of 3.6~Ry is sufficient.
   It gives results as accurate an the nearly all--electron
   projected--augmented wave method (see Ref.~\onlinecite{paw}),
   taking only 1$s^2$ electrons in the core with an energy cutoff
   of 51~Ry.
   The Na$_2$ binding energy and bond--length by these two methods
   differ by less than 3\%.
\end{thebibliography}
\end{document}